\documentclass[aps,prd,twocolumn,groupedaddress,nofootinbib,showpacs]{revtex4}

\usepackage{euscript,graphicx,amssymb,subfigure}

\usepackage{amsfonts}
\usepackage{amsmath}
\usepackage{amssymb}
\usepackage{graphicx}
\usepackage{euscript}
\usepackage{color}
\usepackage{subfigure}

\newcommand{\goodgap}{\hspace{\subfigtopskip} \hspace{\subfigbottomskip}}

\begin{document}

\title{Accelerating $f(T)$ gravity models constrained by recent cosmological data}

\author{Vincenzo F. Cardone}
\email{winnyenodrac@gmail.com}
\affiliation{I.N.A.F.\,-\,Osservatorio Astronomico di Roma, via Frascati 33, 00040 - Monte Porzio Catone (Roma), Italy}

\author{Ninfa Radicella}
\email{ninfa.radicella@gmail.com}
\affiliation{Dipartimento di Fisica ``E.R. Caianiello", Universit\`{a} di Salerno and I.N.F.N. - Sez. di Napoli, GC di Salerno, Via Ponte Don Melilo, 84081 Fisciano (Sa), Italy}

\author{Stefano Camera}
\email{stefano.camera@ist.utl.pt}
\affiliation{CENTRA, Instituto Superior T\'ecnico, Universidade T\'ecnica de Lisboa,
Av. Rovisco Pais 1, 1049\,-\,001 Lisboa, Portugal}

\begin{abstract}

Generalised Teleparallel gravity, also referred to as $f(T)$ gravity, has been recently proposed as an extended theory of gravitation able to give rise to an accelerated expansion in a matter only universe. The cosmic speed up is driven by an effective torsion fluid whose equation of state depend on the $f(T)$ function entering the modified gravity Lagrangian. We focus on two particular choices for $f(T)$ which share the nice property to emulate a phantom divide crossing as suggested by some recent data. We check their viability contrasting the predicted background dynamics to the Hubble diagram as traced by both Type Ia Supernovae (SNeIa) and Gamma Ray Bursts (GRBs), the measurement of the rate expansion $H(z)$, the Baryon Acoustic Oscillations (BAOs) at different redshifts, and the Cosmic Microwave Background Radiation (CMBR) distance priors. Both $f(T)$ models turn out to be in very good agreement with this large dataset so that we also investigate whether it is possible to discriminate among them relying on the different growth factors.

\end{abstract}


\maketitle

\section{Introduction}

The discovery of the acceleration of the universe through the SNeIa Hubble diagram \cite{SNeIaNobel}, recently awarded by the Nobel prize, has been latter confirmed by wide range of data, from more recent SNeIa data to BAOs and CMBR anisotropies \cite{others,WMAP7}. On the other hand, such overwhelming abundance of observational evidences in favour of the cosmic speed up does not fit in the framework of  General Relativity (GR) making clear that out theoretical background is seriously flawed. The naive interpretation that an unexpected new ingredient in the form of a negative pressured component is driving the acceleration poses difficult questions on its nature and nurture and introduces further problems hard to be solved. It has therefore gained more and more attraction the hypothesis that the cosmic speed up is rather the first signal of a breakdown of our understanding of the laws of gravity on cosmological scales. Motivated by this consideration, much attention has been dedicated to $f(R)$ theories \cite{fofR} where the scalar curvature $R$ is replaced by a suitably chosen function $f(R)$ in the gravity Lagrangian.

Looking for a correction to GR, it is instructive to remember that an equivalent formulation is represented by teleparallelism. In this theory, torsion, instead of curvature, is responsible of the gravitational interaction \cite{einstein28,hayashi79} and the Weitzenbock connection replaces the Levi\,-\,Civita one on the underlying Riemann\,-\,Cartan spacetime. In this scenario, gravitational interaction is not replaced by geometry and the torsion acts as a force, allowing the interpretation of gravity as a gauge theory of the translation group \cite{arcos04}. Despite conceptual differences, teleparallel gravity and GR yield equivalent dynamics so that the interpretation of the gravitational interaction in terms of a curved or torsioned spacetime is only a matter of convenience, at least at the classical level.

In as much the same way as for $f(R)$ theories, one can obtain a generalised teleparallel gravity replacing $T$ with a generic function $f(T)$ thus opening the way to a rich phenomenology. A particular important consequence is the breakdown of the equivalence with the classical GR with the two theories now predicting a radically different dynamics \cite{ferraro07}. Modified teleparallel gravity preserves the advantage of giving equations that are still second order in field derivatives opposite to the fourth order equations deduced in $f(R)$ gravity thus avoiding unpleasant pathologies. On the other hand, it suffers from the lack of Local Lorentz Invariance (LLI) so that all the 16 components of the vierbien are independent and one cannot simply fix 6 of them by a gauge choice \cite{li11}.

A critical role in generalised teleparallel theories is played by the choice of the functional expression for $f(T)$. The lack of firmly established theoretical constraints leaves open the way to a wide range of possibilities which can only be validated a posteriori, i.e. by contrasting their predictions with the observational data. This is the aim of the present work where we focus our attention on two particular classes able to give rise to a phantom\,-\,like behaviour of the effective torsion fluid. We then test these two models against SNeIa\,+\,GRB Hubble diagram, $H(z)$ measurements from cosmic chronometers, BAOs data and the CMBR distance priors. Although wide, the present dataset only traces the background expansion so that we will also investigate whether further insight into the properties of these models can be obtained by the analysis of the growth factor being this latter a quick way to look at how perturbations evolve in the proposed modified teleparallel scenarios.

The layout of the paper is as follows. In Sect.\,\ref{f(T)}, we briefly review the cosmology of $f(T)$ gravity and present the two models we are going to analyse. Sect.\,\ref{data} is devoted to the likelihood analysis and the results, while Sect.\,\ref{growth} enlarges the study to the growth factor. Finally, conclusions and perspectives are drawn in Sect.\,\ref{conclusions}.

\section{$f(T)$ gravity}\label{f(T)}

Teleparallelism is a dynamical theory for the vierbein $\{e^a\}$, whose components in a given coordinate basis $e^a_\mu$ help in introducing the metric as a subsidiary field\,:

\begin{displaymath}
g_{\mu \nu}(x) = \eta_{a b} e^a_\mu(x) e^b_\nu(x) \ ,
\end{displaymath}
where $\eta_{a b} = \text{diag}(1,-1,-1,-1)$. The dynamics is then described by the action

\begin{equation}
{\cal{S}} = \frac{1}{16 \pi G} \int{\left [ T + f(T) \right ] e d^4x} + {\cal{S}}_M \ ,
\label{eq: action}
\end{equation}
where $e = \text{det} \  e^a_\mu = \sqrt{-\text{det}(g_{\mu \nu})}$ and ${\cal{S}}_M$ is the action for the matter fields.  In Eq.(\ref{eq: action}), a key role is played by $f(T)$ which is a differentiable function of the torsion tensor $T$ defined as

\begin{displaymath}
T = S^\rho_{\mu \nu} T_\rho ^{\mu \nu}
\end{displaymath}
with

\begin{displaymath}
S^\rho_{\mu \nu} = \frac{1}{4} \left ( T^{\rho}_{\mu \nu} - T_{\mu \nu}^{\rho}+T_{\nu \mu}^{\rho} \right ) +
\frac{1}{2} \delta^\rho_\mu T_{\sigma \nu}^{\sigma} - \frac{1}{2} \delta^\rho_\nu T_{\sigma \mu}^{\sigma} \ ,
\end{displaymath}

\begin{displaymath}
T^\lambda_{\mu \nu} = e^\lambda_a \left( \partial_\nu e^a_\mu - \partial_\mu e^a_\nu \right ) \ .
\end{displaymath}
Varying the action with respect to the vierbein components $e^a_\mu(x)$, one gets the field equations

\begin{eqnarray}
&&e^{-1}\partial_\mu(e\   e_a^\rho S_{\rho}^{\ \mu\nu})(1+f_T)+e_{a}^\lambda S_{\rho}^{\ \nu\mu} T^{\rho}_{\ \mu\lambda} (1+f_T)\nonumber\\
&&+ e^{\rho}_a S_{\rho}^{\ \mu\nu}\partial_\mu (T) f_{TT}+\frac{1}{4}e_a^\nu (T+f) = 4\pi G e_a^\mu {\cal{T}}_\mu^\nu,
\label{eq: fieldeqs}
\end{eqnarray}
where ${\cal{T}}^\nu_\mu$ is the matter energy\,-\,momentum tensor and subscripts $T$ denote differentation with respect to $T$.

One can naively expect that, in order to get the cosmological counterpart of the field equations, one has simply to insert the Robertson\,-\,Walker (RW) metric into Eqs.(\ref{eq: fieldeqs}). Actually, some care is needed since, due to the lack of LLI, two pairs of vierbein that lead to the same metric tensor are not equivalent from the point of view of the theory. Nevertheless, in case of spatially flat RW metric, a convenient choice is represented by the diagonal vierbein, i.e. \cite{cosmof(T)}

\begin{displaymath}
e^0 = dt \ \ , \ \  e^i = a(t) dx^i \ \ ,
\end{displaymath}
where $a(t)$ is the scale factor as function of cosmic time $t$. With such a choice, the dynamical equations become\,:

\begin{equation}
\left \{
\begin{array}{l}
\displaystyle{H^2 = \frac{8 \pi G}{3} \rho - \frac{1}{6} f(T) - 2 H^2 f_T(T)} \\
 \\
\displaystyle{\left ( H^2 \right )^{\prime} = \frac{16 \pi G p + 6 H^2 + f(T) + 12 H^2 f_T(T)}{24 H^2 f_{TT}(T) - 2 - 2f_T(T)}}
\end{array}
\right .  \ ,
\label{eq: ftfried}
\end{equation}
while the torsion scalar reduces to $T = -6 H^2$ with $H = \dot{a}/a$ the usual Hubble parameter and $(\rho, p)$ the energy density and pressure of the matter component. Note that hereafter we will denote with a prime and with a dot differentiation with respect to $\ln{a}$ and $t$, respectively.

The modified Friedmann equations (\ref{eq: ftfried}) can be rewritten in the usual form by introducing an effective dark torsion fluid with energy density reading

\begin{equation}
\rho_T = \frac{2 T f_T(T) - f(T)}{16 \pi G}  \ .
\label{eq: ftrho}
\end{equation}
Since matter still minimally couples to gravity only, its conservation equation will be unaffected so that we still have $\rho_M \propto a^{-3}$ and $\rho_r \propto a^{-4}$ for the scaling laws of dust matter and radiation. Imposing the Bianchi identities and the diffeomorphism invariance of the theory, the conservation equation for the effective torsion fluid reads

\begin{displaymath}
\dot{\rho}_T + 3 H (1 + w_T) \rho_T = 0
\end{displaymath}
having set

\begin{equation}
w_T = -\frac{f/T -f_T+ 2 T f_{TT} + (\Omega_r/3)(f_T + 2 T f_{TT} )}{(1 + f_T + 2 T f_{TT}) (f/T - 2 f_T)}
\label{eq: fteos}
\end{equation}
for the equation of state (eos) of the dark torsion fluid. Note the coupling to the radiation energy density through the term $\Omega_r(a) = 8 \pi G \rho_r(a)/3H^2(a)$. For $f(T) = 0$, one has $\rho_T = 0$ and teleparallel gravity goes back to the standard GR, while the choice $f(T) = const$ gives $w_T = -1$ and the $\Lambda$CDM model is recovered.

Moving away from the naive choice $f(T) = const$ gives rise to rich phenomenology in the dynamical evolution of generalised teleparallel theories. There are actually almost no theoretical hints on the functional form of $f(T)$ with, on the contrary, many possible expressions leading to an accelerated expansion. It is particularly interesting to look at models which are able to give rise to an effective eos (defined later) crossing the phantom divide, i.e., $w_{eff}(z) > -1$ for $z > z_{ph}$ with $z_{ph}$ the phantom divide redshift. Two recently proposed model of this kind can be obtained setting \cite{WY11}

\begin{figure}
\centering
\includegraphics[width=7.5cm]{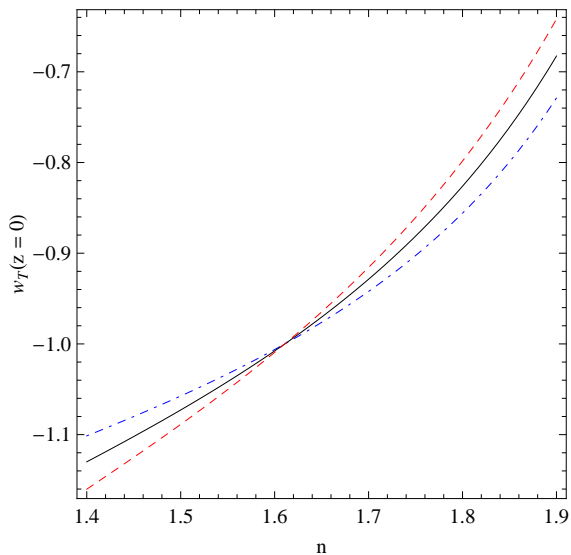}
\caption{Present day value of the torsion eos for the tanh model setting $\Omega_M = 0.20$ (dot dashed blue), $0.25$ (solid black), $0.30$ (dashed red) and the fiducial $\Omega_r$ used in the text.}
\label{fig: wzefftanh}
\end{figure}

\begin{equation}
f(T) = \left \{
\begin{array}{l}
\displaystyle{\alpha (-T)^n \tanh{\left ( \frac{T_0}{T} \right )}} \\
~ \\
\displaystyle{\alpha (-T)^n \left [ 1 - \exp{\left ( - p \frac{T_0}{T} \right )} \right ]} \\
\end{array}
\right . \ ,
\label{eq: ftmods}
\end{equation}
where the subscript $0$ denotes present day quantities.

In the first case, which we will refer to as the tanh model, one should set $n > 3/2$ in order to have a dark torsion fluid with a positive energy density, while the same condition leads to $n > 1/2$ for the second choice, referred to hereafter as the exp model. The constant $\alpha$ can be expressed as a function of the other parameters by inserting Eq.(\ref{eq: ftmods}) into the Friedmann equations (\ref{eq: ftfried}) and imposing $E^2(z = 0) = H^2(z = 0)/H_0^2 = 1$. This gives\,:

\begin{equation}
\alpha = \left \{
\begin{array}{l}
\displaystyle{- \frac{\left ( 6 H_0^2 \right )^{1 - n} (1 - \Omega_M - \Omega_r)}{2 {\rm sech}^2{(E^2 = 1)} + (1 - 2n) \tanh{(E^2 = 1)}}} \\
~ \\
\displaystyle{- \frac{\left ( 6 H_0^2 \right )^{1 - n} (1 - \Omega_M - \Omega_r)}{1 - 2n - (1 - 2n + 2p) {\rm e}^p}} \\
\end{array}
\right . \ ,
\label{eq: alpha}
\end{equation}
for the tanh and exp model. Note that we have dropped the $0$ subscripts for the dust and radiation present day density parameters $(\Omega_M, \Omega_r)$ to simplify the notation.

\begin{figure}
\centering
\includegraphics[width=7.5cm]{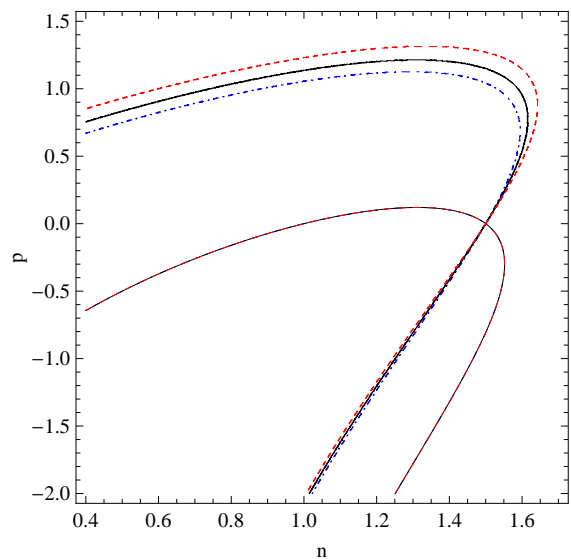}
\caption{Contour plots in the $(n, p)$ plane imposing $w_{T}(z = 0) = -1.0$ for the exp model with $\Omega_M = 0.20$ (dot dashed blue), $0.25$ (solid black), $0.30$ (dashed red) and the fiducial $\Omega_r$ used in the text. Note that the lower set of contours are almost perfectly superimposed.}
\label{fig: wzeffexp}
\end{figure}

It is instructive to look at the effective torsion fluid eos $w_T(z)$ for the two $f(T)$ models we are considering. We do not give here the lengthy expressions which can be obtained by inserting Eqs.(\ref{eq: ftmods}) into the definition (\ref{eq: fteos}), but show how its present day value $w_{T}(z = 0)$ depend on the $f(T)$ parameters in Figs.\,\ref{fig: wzefftanh} and \ref{fig: wzeffexp} for the tanh and exp model, respectively. Since the $\Lambda$CDM model is known to very well fit the data, it is reasonable to expect that, in order to fit the same data, the parameter space of both models will collapse in a region giving $w_{T}(z = 0) = -1$. Fig.\,\ref{fig: wzefftanh} shows that, for the tanh model, this will favour values of $n \simeq 1.6$ and that the larger is $n$, the smaller $\Omega_M$ must be in order to have a today effective eos close to the $\Lambda$CDM one. This same requirement works in a different way for the exp model. Fig.\,\ref{fig: wzeffexp}, indeed, shows that, for given $(\Omega_M, n)$ values, there can be more models with $w_{T}(z = 0) = -1$ but different $p$. Put in other words, the $(n, p)$ parameters are intrinsically degenerate so that the condition $w_{T}(z = 0) = -1$ is unable to discriminate among different parameter sets. Actually, when fitting the data, we do not only look at the present day $w_{eff}$ value, but also to its variation with $z$. As we will see, this helps breaking (at least partially) the degeneracy among $(n, \Omega_M)$ for the tanh model and $(n, p)$ for the exp one. Figs.\,\ref{fig: wzefftanh} and \ref{fig: wzeffexp} are nevertheless useful to quickly grasp what is the region of the parameter space of greater interest for both the tanh and exp $f(T)$ models.

\section{$f(T)$ models vs data}\label{data}

The above considerations on the effective eos suggest that the proposed generalised teleparallel scenarios can represent viable alternatives to the $\Lambda$CDM model. To check whether this is indeed the case, we contrast them with a wide set of cosmological data as described below.

\subsection{Likelihood analysis}

In order to answer the question whether $f(T)$ gravity can reproduce the observed Universe, we will explore the model parameter spaces by investigating the following likelihood function

\begin{equation}
{\cal{L}}({\bf p}) = {\cal{L}}_{\mu}({\bf p}) \ \times \ {\cal{L}}_{H}({\bf p}) \ \times \ {\cal{L}}_{BAO}({\bf p}) \ \times \ {\cal{L}}_{CMB}({\bf p})
\label{eq: totlike}
\end{equation}
where the set of model parameters is

\begin{displaymath}
{\bf p} = \left \{
\begin{array}{l}
(\Omega_M, h, n) \\
~ \\
(\Omega_M, h, n, p) \\
\end{array}
\right .
\end{displaymath}
for the tanh and exp models, respectively, $h$ being the Hubble constant $H_0$ in units of $100 \ {\rm km/s/Mpc}$, and we have set the present day radiation density parameter as

\begin{displaymath}
\Omega_{r} = \omega_{\gamma} h^{-2} (1 + 0.2271 N_{eff})
\end{displaymath}
with $(\omega_{\gamma}, N_{eff}) = (2.469 \times 10^{-5}, 3.04)$ in agreement with WMAP7 \cite{WMAP7} constraints. All the terms in the full likelihood can be written as\,:

\begin{displaymath}
{\cal{L}}_i({\bf p}) = \frac{\exp{\left [ - \chi^2_i({\bf p})/2 \right ]}}{(2 \pi)^{{\cal{N}}_i/2} \Gamma_i^{1/2}}
\end{displaymath}
where $({\cal{N}}_i, \Gamma_i, \chi^2_i)$ depend on the dataset used.

The first term refers to the Hubble diagram, i.e. the distance modulus $\mu$ as function of the redshift $z$. This is related to the underlying cosmological model as

\begin{equation}
\mu(z) = 25 + 5 \log{d_L(z)} = 25 + 5 \log{\left [ (1 + z) r(z) \right ]}
\label{eq: defmu}
\end{equation}
with $d_L(z) = (1 + z) r(z)$ the luminosity distance and

\begin{equation}
r(z) = \frac{c}{H_0} \int_{0}^{z}{\frac{dz^{\prime}}{E(z^{\prime}, {\bf p})}}
\label{eq: defrz}
\end{equation}
the comoving distance. As a tracer, we use both the Union2 SNeIa dataset \cite{Union2}, comprising ${\cal{N}}_{SNeIa} = 557$ objects wiht $0.015 \le z \le 1.40$, and ${\cal{N}}_{GRB} = 64$ GRBs probing the redshift range $(1.48, 5.60)$ with the distance modulus estimated in \cite{Marcy} starting from the data in \cite{XS10}. For both datasets, it is

\begin{displaymath}
\Gamma_i = \prod_{j = 1}^{{\cal{N}}_{i}}{\left ( \sigma_{j}^2 + \sigma_{int}^2 \right )} \ ,
\end{displaymath}

\begin{displaymath}
\chi^2_{i} = \sum_{j = 1}^{{\cal{N}}_{j}}{\left [ \frac{\mu_{obs}(z_j) - \mu_{th}(z_j, {\bf p})}{\sqrt{\sigma_{j}^2 + \sigma_{int}^2}} \right ]^2} \ , \end{displaymath}
where $\sigma_j$ is the measurement error for the $j$\,-\,th object, while $\sigma_{int}$ takes care of the intrinsic scatter of the tracer around the relations used to estimate its $\mu$ value. For SNeIa, we set $\sigma_{int} = 0$, while, for GRBs, this is a nuisance parameter we marginalize over.

While the Hubble diagram probes the integrated expansion rate, the second dataset we use explicitly refers to the Hubble parameter $H(z)$ as determined through the differential age method \cite{JL02}. By using red envelope galaxies as cosmic chronometers \cite{S10II}, Stern et al. have estimated $H(z)$ over the redshift range $0.10 \le z \le 1.75$ \cite{S10I}. To this sample, we add the $H_0$ determination from local distance ladders derived by the SHOES collaboration \cite{SHOES} and define a total $\chi^2$ as for the SNeIa one.

While SNeIa and GRBs probe the distance\,-\,redshift relation as standardizeable candles, Baryon Acoustic Oscillations (BAOs) work as standard rulers. We therefore add the term ${\cal{L}}_{BAO}$ to the full likelihood following the method detailed in \cite{WiggleZ}. To this end, three different sets of data are used. First, one use the 6dFGRS \cite{6dFGRS} and the SDSS \cite{P10} surveys to determine the scaled volume distance parameter

\begin{equation}
d_z = \frac{r_s(z_d)}{d_V(z)} = r_s(z_d) \times \left [ \frac{c z r^2(z)}{H_0 E(z)} \right ]^{-\frac{1}{3}} \ ,
\label{eq: defdz}
\end{equation}
with the sound horizon to distance $z$ given by

\begin{equation}
r_s(z) = \frac{c}{\sqrt{3} H_0} \int_{z}^{\infty}{\frac{E^{-1}(z) dz^{\prime}}{\sqrt{1 + (3 \omega_b)/(4 \omega_r) (1 + z^{\prime})^{-1}}}}
\label{eq: defrs}
\end{equation}
and $z_d$ is the drag redshift. The $d_z$ value at $z = 0.106$ and its error is taken from \cite{6dFGRS}, while \cite{P10} gives $d_z$ for $z = 0.20$ and $z = 0.35$ with the corresponding covariance matrix. Further BAOs constraints come from the WiggleZ survey \cite{WiggleZ} which recommend to use as observable quantity the acoustic parameter \cite{Eis05}

\begin{equation}
{\cal{A}}(z) = \frac{\sqrt{\Omega_M H_0^2} d_V(z)}{c z}
\label{eq: defaz}
\end{equation}
with the volume distance $d_V(z)$ given in Eq.(\ref{eq: defdz}). We use the observed values and their covariance matrix for ${\cal{A}}(z)$ determinations at $z = (0.44, 0.60, 0.70)$ reported in \cite{WiggleZ}. We finally define the BAO likelihood as

\begin{displaymath}
{\cal{L}}_{BAO}({\bf p}) = {\cal{L}}_{6dFGRS}({\bf p}) \ \times \ {\cal{L}}_{SDSS}({\bf p}) \ \times \ {\cal{L}}_{WiggleZ}({\bf p})
\end{displaymath}
with

\begin{displaymath}
{\cal{L}}_{6dFGRS} = \frac{1}{\sqrt{2 \pi \sigma_{0.106}^2}} \ \exp{\left \{ - \frac{1}{2}
\left [ \frac{d_{0.106}^{obs} - d_{0.106}^{th}({\bf p})}{\sigma_{0.106}} \right ]^2 \right \}} \ ,
\end{displaymath}

\begin{eqnarray}
{\cal{L}}_{SDSS} & = & \frac{1}{(2 \pi)^{{\cal{N}}_{SDSS}} | {\bf C}_{SDSS} |^{1/2}} \nonumber \\
~ & \times & \exp{\left [ - \frac{{\bf D}_{SDSS}^{T}({\bf p}) {\bf C}_{SDSS}^{-1} {\bf D}_{SDSS}({\bf p})}{2} \right ]} \nonumber \ ,
\end{eqnarray}

\begin{eqnarray}
{\cal{L}}_{WiggleZ} & = & \frac{1}{(2 \pi)^{{\cal{N}}_{WiggleZ}} | {\bf C}_{WiggleZ} |^{1/2}} \nonumber \\
~ & \times & \exp{\left [ - \frac{{\bf D}_{WiggleZ}^{T}({\bf p}) {\bf C}_{WiggleZ}^{-1} {\bf D}_{WiggleZ}({\bf p})}{2} \right ]} \nonumber \ ,
\end{eqnarray}
where ${\bf D}$ is a ${\cal{N}}_i$ dimensional vector with the difference between observed and predicted values and ${\bf C}_i$ the corresponding covariance matrix.

\begin{table}
\begin{center}
\begin{tabular}{cccccc}
\hline
Id & $x_{BF}$ & $\langle x \rangle$ & $\tilde{x}$ & $68\% \ {\rm CL}$ & $95\% \ {\rm CL}$ \\
\hline \hline
~ & ~ & ~ & ~ & ~ & ~ \\
$\Omega_M$ & 0.286 & 0.286 & 0.287 & (0.274, 0.299) & (0.264, 0.311) \\
~ & ~ & ~ & ~ & ~ & ~ \\
$h$ & 0.719 & 0.722 & 0.722 & (0.712, 0.734) & (0.702, 0.745) \\
~ & ~ & ~ & ~ & ~ & ~ \\
$n$ & 1.616 & 1.610 & 1.615 & (1.581, 1.636) & (1.547, 1.667) \\
~ & ~ & ~ & ~ & ~ & ~ \\
\hline
\end{tabular}
\caption{Constraints on the parameters for the tanh model. Columns are as follows\,: 1.) id parameter, 2.) best fit, 3.) mean, 4.) median, 5.), 6.) $68\%$ and $95\%$ confidence levels.}
\label{tab: tanhfit}
\end{center}
\end{table}

The last term in the likelihood finally refers to the WMAP7 distance priors which have been recommended as a quick and efficient way to include the CMBR constraints without computing the full anisotropy spectrum. Following \cite{WMAP7}, we then define the CMBR likelihood similar to the SDSS and WiggleZ ones above, but now the observable quantities are the redshift $z_{\star}$ to the last scattering surface, computed using the approximated formulae in \cite{HS96}, the acoustic scale $\ell_A = \pi r(z_{\star})/r_s(z_{\star})$ and the shift parameter ${\cal{R}}$ \cite{ShiftPar}

\begin{equation}
{\cal{R}} = \frac{\sqrt{\Omega_M} r(z_{\star})}{c/H_0} \ .
\label{eq: defshiftpar}
\end{equation}
Note that, in order to use the distance priors, one should implicitly assume that the early universe is matter dominated with no significant contribution from dark energy, either as an actual fluid or an effective one induced by nonlinear terms in the gravity Lagrangian. Although the additional $f(T)$ term does not vanish at high $z$ for all the possible combinations of the model parameters, it is nevertheless easy to check that the effective torsion fluid energy density is by orders of magnitude smaller than the matter one at $z_{\star}$ so that we can still rely on the WMAP7 determination of the distance priors.

\begin{figure*}
\centering
\subfigure{\includegraphics[width=7.5cm]{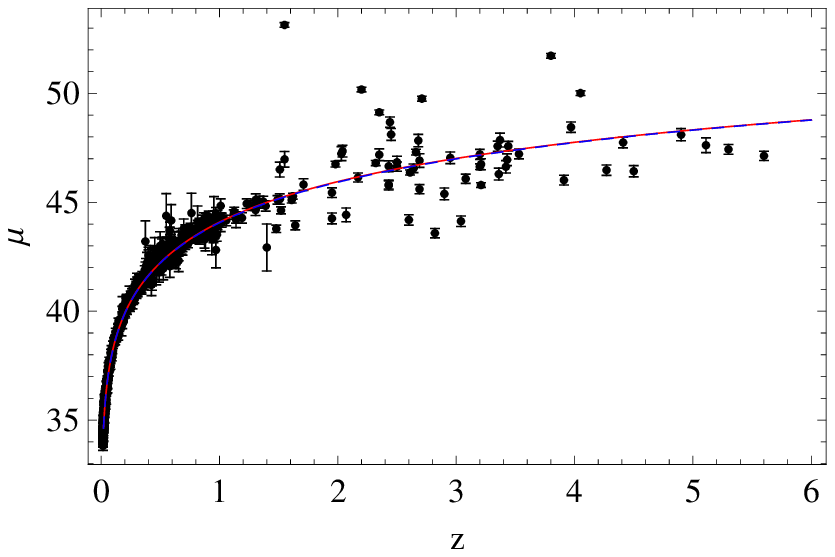}} \goodgap
\subfigure{\includegraphics[width=7.5cm]{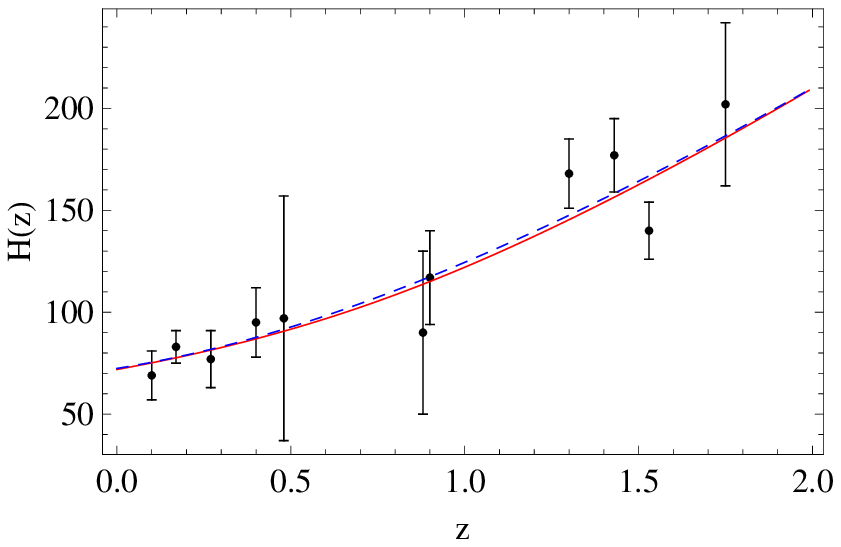}}
\caption{Best fit curves superimposed to the SNeIa\,+\,GRB Hubble diagram (left) and $H(z)$ data (right) for the tanh (red solid) and exp (blue dashed) models. Note that the $\mu(z)$ curves are almost perfectly superimposed so no difference is seen in the plot.}
\label{fig: bfplot}
\end{figure*}

In order to efficiently explore the parameter space, we use a Markov Chain Monte Carlo (MCMC) code running multiple chains and checking the convergence through the Gelman\,-\,Rubin criterium \cite{GR92}. The best fit parameters will be the ones maximizing the full likelihood, but the most reliable constraints on each single parameter $p_i$ are obtained by marginalizing over all the parameters but the $i$\,-\,th one. We will give the mean and median as reference value and use the $68$ and $95\%$ confidence ranges as our $1$ and $2 \sigma$ errors (although this is formally correct only if the marginalized likelihood is a Gaussian function).

\begin{table}
\begin{center}
\begin{tabular}{cccccc}
\hline
Id & $x_{BF}$ & $\langle x \rangle$ & $\tilde{x}$ & $68\% \ {\rm CL}$ & $95\% \ {\rm CL}$ \\
\hline \hline
~ & ~ & ~ & ~ & ~ & ~ \\
$\Omega_M$ & 0.284 & 0.286 & 0.287 & (0.276, 0.297) & (0.265, 0.308) \\
~ & ~ & ~ & ~ & ~ & ~ \\
$h$ & 0.724 & 0.731 & 0.731 & (0.723, 0.740) & (0.713, 0.749) \\
~ & ~ & ~ & ~ & ~ & ~ \\
$n$ & 1.152 & 0.757 & 0.736 & (0.577, 0.939) & (0.514, 1.103) \\
~ & ~ & ~ & ~ & ~ & ~ \\
$p$ & 0.814 & -0.110 & -0.100 & (-0.263, 0.046) & (-0.395, 0.131) \\
~ & ~ & ~ & ~ & ~ & ~ \\
\hline
\end{tabular}
\caption{Same as Table \ref{tab: tanhfit} but for the exp model.}
\label{tab: expfit}
\end{center}
\end{table}

\subsection{Results}

The best fit parameters and marginalized constraints for the tanh and exp model are given in Tables \ref{tab: tanhfit} and \ref{tab: expfit}, while Fig.\,\ref{fig: bfplot} shows the remarkable agreement among the best fit models predictions and the SNeIa\,+\,GRB Hubble diagram and $H(z)$ data. The overall good quality of the fit may be further appreciated by comparing the model predictions for the BAO and CMB quantities with the observed values. For the tanh model, we get

\begin{displaymath}
\left \{
\begin{array}{lll}
d_{0.106}^{bf} = 0.3418 & {\rm vs} & d_{0.106}^{obs} = 0.336 \pm 0.015 \\
~ & ~ & ~ \\
d_{0.200}^{bf} = 0.1864 & {\rm vs} & d_{0.200}^{obs} = 0.1905 \pm 0.0061 \\
~ & ~ & ~ \\
d_{0.350}^{bf} = 0.1173 & {\rm vs} & d_{0.350}^{obs} = 0.1097 \pm 0.0036 \\
\end{array}
\right . \ ,
\end{displaymath}

\begin{displaymath}
\left \{
\begin{array}{lll}
{\cal{A}}^{bf}(0.44) = 0.467 & {\rm vs} & {\cal{A}}^{obs}(0.44) = 0.474 \pm 0.034 \\
~ & ~ & ~ \\
{\cal{A}}^{bf}(0.60) = 0.442 & {\rm vs} & {\cal{A}}^{obs}(0.60) = 0.442 \pm 0.020 \\
~ & ~ & ~ \\
{\cal{A}}^{bf}(0.73) = 0.422 & {\rm vs} & {\cal{A}}^{obs}(0.73) = 0.424 \pm 0.021 \\
\end{array}
\right . \ ,
\end{displaymath}

\begin{displaymath}
\left \{
\begin{array}{lll}
\ell_A^{bf} = 302.66 & {\rm vs} & \ell_A^{obs} = 302.09 \pm 0.76 \\
~ & ~ & ~ \\
{\cal{R}}^{bf} = 1.733 & {\rm vs} & {\cal{R}}^{obs} = 1.725 \pm 0.018 \\
~ & ~ & ~ \\
z_{\star}^{bf} = 1092.04 & {\rm vs} & z_{\star}^{obs} = 1091.30 \pm 0.91 \\
\end{array}
\right . \ ,
\end{displaymath}
so that the best fit tanh model predictions are well within $1 \sigma$ from the observed values. For the exp model, we get

\begin{displaymath}
\left \{
\begin{array}{lll}
d_{0.106}^{bf} = 0.3428 & {\rm vs} & d_{0.106}^{obs} = 0.336 \pm 0.015 \\
~ & ~ & ~ \\
d_{0.200}^{bf} = 0.1865 & {\rm vs} & d_{0.200}^{obs} = 0.1905 \pm 0.0061 \\
~ & ~ & ~ \\
d_{0.350}^{bf} = 0.1121 & {\rm vs} & d_{0.350}^{obs} = 0.1097 \pm 0.0036 \\
\end{array}
\right . \ ,
\end{displaymath}

\begin{displaymath}
\left \{
\begin{array}{lll}
{\cal{A}}^{bf}(0.44) = 0.465 & {\rm vs} & {\cal{A}}^{obs}(0.44) = 0.474 \pm 0.034 \\
~ & ~ & ~ \\
{\cal{A}}^{bf}(0.60) = 0.439 & {\rm vs} & {\cal{A}}^{obs}(0.60) = 0.442 \pm 0.020 \\
~ & ~ & ~ \\
{\cal{A}}^{bf}(0.73) = 0.418 & {\rm vs} & {\cal{A}}^{obs}(0.73) = 0.424 \pm 0.021 \\
\end{array}
\right . \ ,
\end{displaymath}

\begin{displaymath}
\left \{
\begin{array}{lll}
\ell_A^{bf} = 302.54 & {\rm vs} & \ell_A^{obs} = 302.09 \pm 0.76 \\
~ & ~ & ~ \\
{\cal{R}}^{bf} = 1.735 & {\rm vs} & {\cal{R}}^{obs} = 1.725 \pm 0.018 \\
~ & ~ & ~ \\
z_{\star}^{bf} = 1092.12 & {\rm vs} & z_{\star}^{obs} = 1091.30 \pm 0.91 \\
\end{array}
\right . \ ,
\end{displaymath}
still in good agreement with the BAO and CMBR data.

\begin{figure*}
\centering
\subfigure{\includegraphics[width=7.5cm]{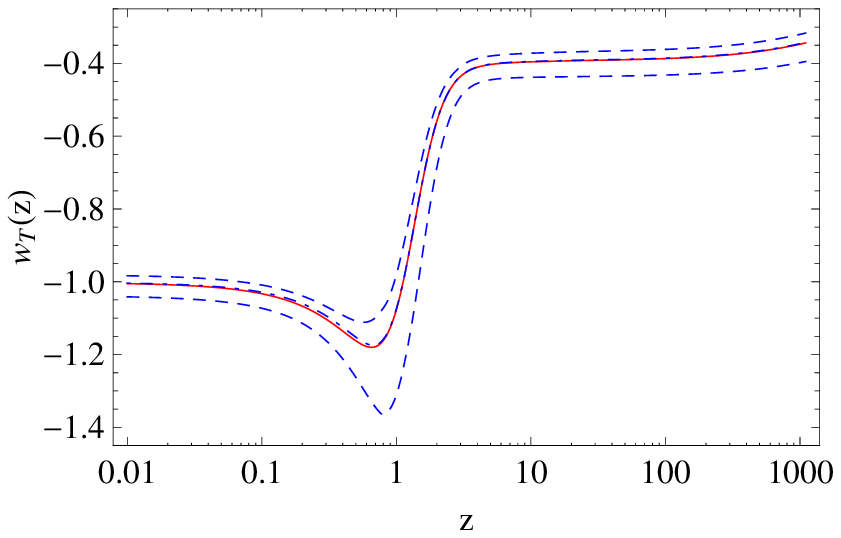}} \goodgap
\subfigure{\includegraphics[width=7.5cm]{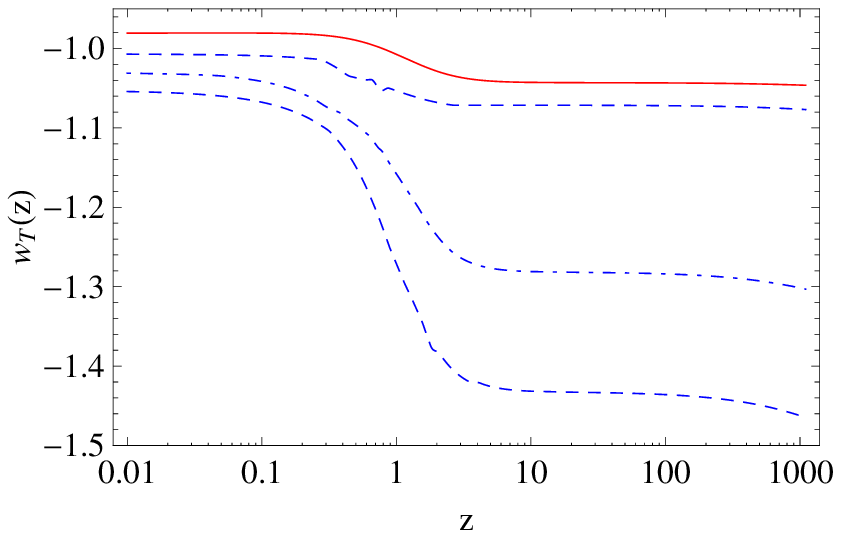}}
\caption{Constrained effective torsion fluid eos $w_T(z)$ for the tanh (left) and exp (right) $f(T)$ models.}
\label{fig: eosplot}
\end{figure*}

The left panel in Fig.\,\ref{fig: bfplot} is quite instructive showing us how the two best fit models can not be discriminated at all based on the Hubble diagram data. This can be easily understood looking at the right panel where one can see that the rate expansion $H(z)$ is quite similar among the two models with the largest deviations occurring in the redshift range $(0.5, 1.5)$. Being of order only few $\%$, it is easy to understand that the integration needed to compute the distance modulus easily washes out the differences between the two models thus leading to the almost perfectly superimposed curves shown in Fig.\,\ref{fig: bfplot}.

Another way to explain why the two models make so similar predictions is to look at the reconstructed effective torsion fluid eos. Fig.\,\ref{fig: eosplot} shows $w_T(z)$ as constrained by the likelihood analysis. In each panel, the red solid curve refers to the best fit model, while the blue dot\,-\,dashed (dashed) one gives, for each $z$, the median value (the $68\%$ confidence range) as estimated evaluating $w_T(z, {\bf p})$ for the model parameters in the MCMC chain. For both best fit models (i.e., look at the red curves), the eos stays close to the cosmological constant value $w = -1$ up to $z \sim 1$ so that the dynamics is approximately the same as the $\Lambda$CDM one. On the other hand, $w_T(z)$ has a radical different behaviour for the two models in the higher redshift regime. Indeed, while the tanh effective torsion fluid has an increasing eos, the exp model eos stays roughly constant to a value slightly smaller than the present day one thus better mimicking a $\Lambda$CDM scenario. Nevertheless, both best fit models make similar predictions for the CMBR distance priors as a result of the very small contribution of the torsion fluid to the energy budget at the last scattering surface.

As a final remark, it is worth spending some more words on an unusual feature of the exp model. As one can see from Table\,\ref{tab: expfit}, the best fit $(n, p)$ parameters are strongly different from their median values with $p$ being outside the $95\%$ confidence range. As a consequence, the reconstructed torsion fluid eos for the best fit model is outside the $68\%$ confidence range delimited by the blue dashed lines in the right panel of Fig.\,\ref{fig: eosplot}. Such an unexpected result is actually a consequence of the strong degeneracy among the model parameters. As already highlighted when commenting Fig.\,\ref{fig: wzeffexp}, for a given $n$ value, there are more than one $p$ value giving the same $w_{T}(z)$. This intrinsic degeneracy can only partially be broken by fitting data at different redshifts. However, since most part of the data probes the range $z < 1$, all the couples $(n, p)$ giving $w_T(z) \sim -1$ over this range are allowed by the fit. This is indeed the case for the exp models with $(n, p)$ set to their best fit values (leading to $w_T(z = 0) = -1.042$) or to the median ones (giving $w_T(z = 0) = -1.038$). For larger $z$, the median and best fit models eos are radically different, but the predicted distance priors values are nevertheless almost the same since the torsion fluid becomes subdominant so that the details of its eos turn out to be meaningless. In the following, we will however set the $(n, p)$ parameters to their median values since such a choice more reliably takes into account how the data constrain the exp model parameters space.

\section{The growth factor}\label{growth}

\begin{figure}
\centering
\includegraphics[width=7.5cm]{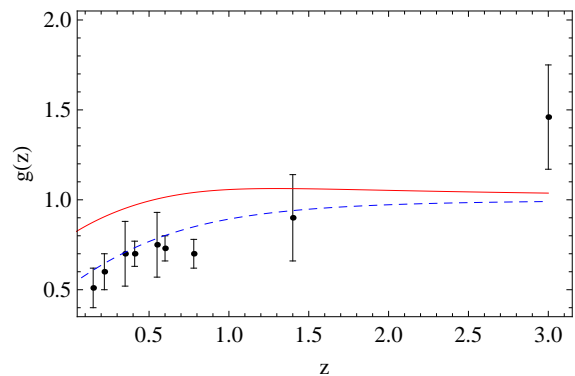}
\caption{Growth factor $g(z)$ for the tanh (red solid) and exp (blue dashed) $f(T)$ models superimposed to the \cite{NP08,C11} data.}
\label{fig: gzplot}
\end{figure}

The above analysis, while confirming that both $f(T)$ models nicely work at reproducing the available data, has also highlighted that probing only the background expansion does not allow us to conclusively discriminate between them and the $\Lambda$CDM scenario. In order to deepen the investigation and look for a way to put stronger constraints, one has to resort to the analysis of the evolution of the perturbations. As a first step, we consider here the growth factor $g = d\ln{\delta}/d\ln{a}$ with $\delta = \delta \rho_M/\rho_M$ the matter density contrast. The $\delta$ evolution can easily be determined solving \cite{ZH11}

\begin{equation}
\ddot{\delta}_M + 2 H \dot{\delta}_M - 4 \pi G_{eff} \rho_M \delta_M = 0 \ ,
\label{eq: growthfac}
\end{equation}
which is formally identical to the $\Lambda$CDM one provided the Newton constant $G$ is replaced by the effective one $G_{eff} = G/(1 + f_T)$. It is worth stressing that Eq.(\ref{eq: growthfac}) actually holds only in the sub\,-\,horizon limit, i.e., under the approximation $k >> {\cal{H}}, {\cal{H}}^{\prime}/{\cal{H}}, {\cal{H}}^2$ with ${\cal{H}}$ the comoving expansion rate. A much more complicated expression \cite{LSB11} holds in the general case and must be used in evaluating large scale probes as the matter power spectrum and, for instance, weak lensing observables. However, for this preliminary investigation, we will use the simpler Eq.(\ref{eq: growthfac}) which is well suited for galactic and galaxy cluster scales.

Fig.\,\ref{fig: gzplot} shows the growth factor for the best fit tanh and the median exp models together with a compilation of $g(z)$ measurement taken from \cite{NP08,C11}. As it is evident, the two models can be easily differentiated from the different $g(z)$ behaviour, while their background expansion is almost the same as shown by the very close agreement between the corresponding $\mu(z)$ and $H(z)$ curves plotted in Fig.\,\ref{fig: bfplot}. According to the $g(z)$ values, structures growth is faster in the tanh than in the exp model so that an observational determination of $g(z)$ can easily discriminate among the two $f(T)$ scenarios. Indeed, using the data plotted, we get $\tilde{\chi}^2 = 15.3$ for the tanh model indicating a strong disagreement, while better results are obtained for the exp model leading to $\tilde{\chi}^2 = 2.1$ (still large, but mainly due to the highest $z$ point which is not even fitted by the $\Lambda$CDM growth factor). These values have, however, to be taken with great caution and should not be considered as a definitive evidence of the exp model working better than the tanh one. First, most of the measurements of $g(z)$ reported in \cite{NP08} actually refers to $g(z)/b(z)$ with $b(z)$ the bias of the galaxy population used to trace the growth factor. Since the bias is related to the underlying gravity theory, a modified Lagrangian can in principle introduce deviations from the standard collapse scenario thus leading to a different and possibly scale dependent bias. Second, $g(z)$ is sometimes obtained from a preliminary modelling of the matter power spectrum as in \cite{C11} which partially relies on the assumption of a fiducial $\Lambda$CDM model to convert from the redshift to the real space clustering. While the background expansion is likely the same as the $\Lambda$CDM one up to intermediate $z$, the modeling of the power spectrum and its distortions can be different so that the inferred $g(z)$ estimate should be taken {\it cum grano salis}. Motivated by these considerations, we prefer to be conservative and consider Fig.\,\ref{fig: gzplot} as a strong evidence that accelerating $f(T)$ models with a quite similar background expansion can be easily discriminated by their growth factor evolution.

\section{Conclusions}\label{conclusions}

Interpreting gravitational interactions in terms of the torsion rather than the scalar curvature leads to the equivalent teleparallel formulation of GR. Adding a further $f(T)$ term into the gravity Lagrangian then modifies General Relativity allowing for a wider range of possible dynamical behaviour. In particular, for some particular choices of the $f(T)$ functional expression, it is possible to get an accelerating expansion in a matter only universe with the effective torsion fluid playing the role of dark energy. Motivated by these considerations, we have here considered two generalisations of teleparallel models assigned by the expression (\ref{eq: ftmods}) for the $f(T)$ function. We have convincingly shown that both models are in excellent agreement with a wide set of cosmological data, from the SNeIa\,+\,GRB Hubble diagram to the $H(z)$ measurements, BAOs and CMBR distance priors.

Having been designed to give an accelerated expansion, the parameter space of both models collapses into a region giving rise to a background dynamics similar to the $\Lambda$CDM one. As a consequence, the two models can be hardly discriminated based on the dataset we have used. It is worth stressing that this is not a limitation of the data, but rather an intrinsic feature of how the models have been worked out. As such, improving the precision of the measurements or increasing the statistics does not help in discriminating among the $f(T)$ models and the $\Lambda$CDM one. On the contrary, one has to resort to different tracers which are related to the evolution of the perturbations, the simplest one being the growth factor $g(z)$. On small scales, the impact of torsion only introduces a redshift dependent rescaling of the gravitational constant which now becomes $G_{eff} = G/(1 + f_T)$. This only modification considerably changes the growth factor leading to a scaling of $g$ with $z$ which strongly depend on the $f(T)$ model considered. As Fig.\,\ref{fig: gzplot} clearly shows, it is easy to discriminate between the two models provided one has a model independent estimate of the growth factor as function of $z$. However, to this end, one should carefully take into account how the collapse of structures take place in the modified regime introduced by the effective torsion field. To this end, the weak field limit of $f(T)$ models must be investigated thus also leading to the determination of the gravitational potential on galactic scales and its impact on the galaxy dynamics.

On the largest scales, the torsion field modifies the growth of perturbations by altering both the gravitational constant and the friction term \cite{C11}. As a consequence a richer phenomenology is achieved possibly leading to other ways to discriminate among $f(T)$ models and dark energy ones. Two candidate probes are the matter power spectrum $P(k)$ and cosmic shear. In the first case, the derivation of $P(k)$ is likely quite easy to achieve the main deviations from the standard GR one being encoded by the scale dependent growth factor. However, the comparison with the data asks for two further effects to be taken into account. First, what we actually observe is the galaxy power spectrum which is related to the matter one through the bias $b$ (if we neglect redshift distortions). While for GR models one can assume $b$ to be scale independent, this is not guaranteed a priori for $f(T)$ models. As a possible way out, one could rely on a parameterized phenomenological ansatz although this strategy could introduce a degeneracy with the $f(T)$ parameters. On smaller scales, where $g(z)$ is scale independent, nonlinear effects start becoming important so that one should find a way to take them into account. As far as we know it, a mapping from the linear $P(k)$ to the nonlinear one is still unavailable for $f(T)$ theories thus preventing a straightforward comparison between predicted and observed $P(k)$. A somewhat better situation holds for the computation of the cosmic shear power spectrum. Indeed, it probes the full matter distribution along the line of sight and not the clustered component. As such, the bias plays no role at all so that there is no need to preliminary investigate how galaxy formation takes place in $f(T)$ models. Moreover, the cosmic shear power spectrum allows us to probe the largest scales where nonlinear effects may be neglected thus putting more emphasis on the modifications due to the deviations of $f(T)$ theory from GR. We therefore consider cosmic shear as a more promising tool to discriminate among rival teleparallel scenarios and between $f(T)$ and GR and deserve a detailed analysis to a forthcoming publication \cite{noipoi}.

\section*{Acknowledgements}

VFC is funded by ASI (Agenzia Spaziale Italiana). NR has been partially supported by a INFN/MICINN collaboration. SC acknowledges support from FCT\,-\,Portugal under grant PTDC/FIS/100170/2008.

\end{document}